\documentclass[%
 aip,
 amsmath,amssymb,
 preprint,%
]{revtex4-1}
\usepackage{graphicx}
\usepackage{dcolumn}
\usepackage{bm}

\usepackage[utf8]{inputenc}
\usepackage[T1]{fontenc}
\usepackage{mathptmx}
\usepackage{romannum}

\usepackage{hyperref}
\usepackage{color}
\allowdisplaybreaks

\makeatletter
\newcommand{\erf}{\mathrm{erf}}
\newcommand{\erfc}{\mathrm{erfc}}
\makeatother

\begin{document}

\author{Qiming Sun}
\email{osirpt.sun@gmail.com}
\affiliation{Quantum Engine LLC, WA 98516, US}

\title{Exact Exchange with Range-separated Algorithm for Thermodynamic Limit of Periodic Hartree-Fock Theory}

\date{\today}

\begin{abstract}
The expensive cost of computing exact exchange in periodic systems
limits the application range of density functional theory with hybrid
functionals.
To reduce the computational cost of exact change, we present a range-separated
algorithm to compute electron repulsion integrals for Gaussian-type crystal
basis.
The algorithm splits the full-range Coulomb interactions into short-range and
long-range parts, which are respectively computed in real and reciprocal space.
This approach significantly reduces the overall computational cost as integrals
can be efficiently computed in both regions.
The algorithm can efficiently
handle large numbers of $\mathbf{k}$ points with limited CPU and memory
resources. As a demonstration, we performed an all-electron $\mathbf{k}$-point
Hartree-Fock calculation for LiH crystal with 1 million Gaussian basis
functions, which was completed on a desktop computer in 1400 CPU hours.
\end{abstract}

\maketitle

\section{Introduction}

Density functional theory (DFT) is the most widely used tool for modeling systems
with periodic boundary conditions (PBC).
Large scale DFT simulations without Hartree-Fock exchange (HFX) were
available for PBC systems of millions of basis functions\cite{Bowler2010,VandeVondele2012}.
The use of hybrid functionals in DFT, which includes HFX contributions in the
DFT exchange-correlation functional, can significantly improve the accuracy of
DFT and widen the range of applications.
However, accurately and efficiently evaluating HFX is a major challenge for PBC DFT calculations.
Problems like numerical divergence and singularity in
$\mathbf{k}$-point sampling may be found in HFX with PBC and various treatments
were developed to overcome these
problems\cite{Pisani1980,Dovesi1983,Gygi1986,Heyd2004,Paier2005,Izmaylov2006,Spencer2008,Guidon2009,Sundararaman2013,Marzari1997,Gygi2009,Marzari2012,Gygi2013,DiStasio2014}.
The concern of using hybrid functional DFT in PBC application is mainly the
computational cost of HFX which can be several orders of magnitude higher then
the rest part.

In crystal simulations, planewave (PW) basis is a natural choice due to its periodicity.
Computing HFX with PW basis\cite{Goerling1996,Chawla1998,Wu2009,Lin2016,Hu2017,Barnes2017}
is extremely demanding and generally considered impractical.
Due to the locality of Gaussian functions, typically less Gaussian functions are
required to describe the atomic characters in a system. It is
more affordable to computing HFX in Gaussian basis than in PW basis.
Gaussian-type crystal basis are widely adopted in many program
packages\cite{Kuehne2020,Dovesi2020,Sun2020,Blum2009,Balasubramani2020,VandeVondele2005}.
For PBC calculations, Gaussian and planewave method\cite{Lippert1997,Alavi2009} (GPW) is a
commonly applied method to compute electron repulsion integrals for Gaussian basis.
In GPW method, PW basis serves as the auxiliary basis of density fitting method.
Discrete Fourier transform is often employed to evaluate the density and
Coulomb potential on PW basis, which has $\mathcal{O}(N_G \log(N_G))$ complexity
for $N_G$ PW basis (or the number of integration grids).
For a calculation of $N_{A}$ atoms and $N_k$ $\mathbf{k}$ points, the
cost of HFX with the GPW method scales as at least $N_{A}^2N_k^2 N_G\log(N_G)$.

The GPW method is not suitable for all-electron calculations as it requires a
high energy cutoff for accurate description of core orbitals, which leads to a
large number of auxiliary PW basis. To reduce the energy cutoff,
pseudopotentials\cite{Hartwigsen1998,Willand2013,Gao2018,Prandini2018,Lu2019} (PP)
become critical in PW basis methods as well as the GPW method.
Although PPs are commonly used in crystal calculations, there are several
issues for applications with PP.
Accuracy and reliability of DFT calculations are badly affected by the quality of PP.
Results obtained from different PPs are often not comparable\cite{Lejaeghere2016}.
Applications with PP are limited to valence properties only. It is impossible to study
properties of core electrons with PP.
As such, all-electron methods receive significant
attentions\cite{Krack2000,Kucukbenli2014,Levchenko2015,Biktagirov2018,Yeh2022,Gygi2023}
since they can naturally address all of these issues.

Gaussian and augmented plane wave (GAPW) method\cite{Lippert1999,Kresse1999,Krack2000} was
developed to address the energy cutoff issue in GPW.
This method uses Gaussian functions to describe electron density in the core
region and calls the GPW method for the valence region only.
However, GAPW cannot be used to compute HFX.
Gaussian density fitting (GDF) were considered\cite{Towler1996,Schwegler1999,Heyd2003,Tymczak2005,Izmaylov2006,Maschio2007,Varga2008,Maschio2008,Guidon2009,Ben2013,Yamada2013,Sun2017,Irmler2018,Wang2020,Lee2022,Sharma2022}
due to its great success for HFX and integral evaluation in molecular systems.
It should be noted that if integrals in GDF are computed with the GPW algorithm,
GDF method may still suffer from the energy cutoff issue in all-electron calculations.
GDF algorithms specialized on all-electron methods were
developed to solve the energy cutoff issue\cite{Sun2017,Ye2021,Ye2021a,Bintrim2022}.

Despite that GPW and GDF methods provide the possibility to compute HFX with PBC
Gaussian basis, their computational requirements of HFX still make them
impractical when targeting the thermodynamic limit in periodic systems.
The GDF method requires a storage of $\mathcal{O}(N_{AO}^3N_k^2)$ for the
intermediate tensors and $\mathcal{O}(N_{AO}^4N_k^2)$ FLOPs for tensor contraction in HFX matrix.
As a result, the feasible problem size of k-point hybrid DFT calculations with
GDF methods is limited to around 100k basis functions.

In this work, we present a range-separated algorithm for the computation
electron repulsion integrals with Gaussian crystal basis.
This algorithm enables us to compute HFX for large periodic systems.
Formulas, possible variants, and integral screening schemes involved in this algorithm
are discussed in Section \ref{sec:theory}.
In Section \ref{sec:results}, we demonstrate the efficiency and accuracy of this
algorithm by performing an all-electron HF calculation for LiH crystal with a million basis functions.

\section{Theory}
\label{sec:theory}
\subsection{Fock matrix with range-separated algorithm}
We consider here a periodic system which consists of repeated unit cells
in terms of the translation vector $\mathbf{m}$ and lattice vectors $\mathbf{a}$
\begin{equation}
  \mathbf{m} = m_x \mathbf{a}_x + m_y \mathbf{a}_y + m_z \mathbf{a}_z.
\end{equation}
Crystal orbitals for a crystal of $N\rightarrow\infty$ unit cells can be
expanded with a set of Bloch orbitals
\begin{equation}
  \phi_\mu(\mathbf{r})^{\mathbf{k}}
  = \frac{1}{\sqrt{N}}\sum_{\mathbf{m}} e^{i\mathbf{k}\cdot \mathbf{m}} \mu^{\mathbf{m}}(\mathbf{r})
\end{equation}
where $\mu^{\mathbf{m}}$ is an real Gaussian-type orbital centered at
$\mathbf{R}_\mu$ in cell $\mathbf{m}$
\begin{equation}
  \mu^{\mathbf{m}}(\mathbf{r}) = \mu(\mathbf{r} - \mathbf{R}_\mu - \mathbf{m}).
\end{equation}

Solving the periodic HF model can lead to the
$\mathbf{k}$-point adapted Roothaan equations
\begin{gather}
  \mathbf{F}^\mathbf{k} \mathbf{C}^\mathbf{k}
  = \mathbf{S}^\mathbf{k} \mathbf{C}^\mathbf{k} \epsilon^\mathbf{k},
\end{gather}
The Fock matrix and overlap matrix at $\mathbf{k}$ are
\begin{gather}
  \mathbf{F}^\mathbf{k} = \mathbf{T}^\mathbf{k} + \mathbf{V}_\mathrm{Nuc}^{\mathbf{k}}
  + \mathbf{J}^\mathbf{k} - \mathbf{K}^{\mathbf{k}},
  \\
  S_{\mu\nu}^{\mathbf{k}}
  = \langle \phi_\mu^{\mathbf{k}}|\phi_\nu^{\mathbf{k}}\rangle.
\end{gather}
The overlap matrix can be computed in terms of the lattice sum over overlap
integrals in real space
\begin{gather}
  S_{\mu\nu}^{\mathbf{k}}
  = \sum_{\mathbf{m}} e^{i\mathbf{k}\cdot\mathbf{m}} S_{\mu^0\nu^\mathbf{m}},
  \label{eq:overlap}
  \\
  S_{\mu^0\nu^m} = \langle \mu^\mathbf{0}|\nu^\mathbf{m}\rangle.
\end{gather}
In a similarly way, we can get the kinetic matrix in $\mathbf{F}^\mathbf{k}$
\begin{gather}
  T_{\mu\nu}^{\mathbf{k}}
  = \sum_{\mathbf{m}} e^{i\mathbf{k}\cdot\mathbf{m}} T_{\mu^0\nu^m}.
\end{gather}
When computing the nuclear attraction matrix $\mathbf{V}_\mathrm{Nuc}^\mathbf{k}$,
we use model charge to screen the long-range interactions then compute
separately the long-range and short-range parts of nuclear attraction integrals.
Relevant formulas have been documented in \onlinecite{Sun2017}.

Given $\mathbf{k}$ denoted density matrices $D^{\mathbf{k}}$,
the HF Coulomb and exchange matrices in $\mathbf{F}^{\mathbf{k}}$ are
\begin{gather}
  J_{\mu\nu}^\mathbf{k}
  =\frac{1}{N_k}\sum_{\mathbf{k}'}\sum_{\kappa\lambda} D_{\lambda\kappa}^{\mathbf{k}'}
  (\phi_\mu^{\mathbf{k}} \phi_\nu^{\mathbf{k}}|\phi_\kappa^{\mathbf{k}'} \phi_\lambda^{\mathbf{k}'}),
  \label{eq:coulomb}
  \\
  K_{\mu\lambda}^\mathbf{k}
  =\frac{1}{N_k}\sum_{\mathbf{k}'}\sum_{\kappa\nu} D_{\nu\kappa}^{\mathbf{k}'}
  (\phi_\mu^{\mathbf{k}} \phi_\nu^{\mathbf{k}'}|\phi_\kappa^{\mathbf{k}'} \phi_\lambda^{\mathbf{k}}).
  \label{eq:exchange}
\end{gather}
The four-center two-electron repulsion integrals (ERI) in above equations are
\begin{gather}
  (\phi_\mu^{\mathbf{k}_\mu} \phi_\nu^{\mathbf{k}_\nu}|\phi_\kappa^{\mathbf{k}_\kappa} \phi_\lambda^{\mathbf{k}_\lambda})
  = \iint \phi_\mu^{\mathbf{k}_\mu}(\mathbf{r}_1)\phi_\nu^{\mathbf{k}_\nu}(\mathbf{r}_1) g(\mathbf{r}_{12})
  \phi_\kappa^{\mathbf{k}_\kappa}(\mathbf{r}_2) \phi_\lambda^{\mathbf{k}_\lambda}(\mathbf{r}_2)
  d^3 \mathbf{r}_1 d^3 \mathbf{r}_2.
  \label{eq:eri}
\end{gather}
This integral can be evaluated in reciprocal space in two steps\cite{Lippert1997,Alavi2009}:
\begin{enumerate}
  \item Fourier transform (FT) the orbital product $\rho_{\mu\nu}$ against the
    PW basis $|\mathbf{G}\rangle = \frac{(2\pi)^{3/2}}{\sqrt{N}}e^{i\mathbf{G}\cdot\mathbf{r}}$
\begin{align}
  \rho_{\mu\nu}^{\mathbf{k}\mathbf{k}'}(\mathbf{G})
  &= \int e^{-i\mathbf{G}_{\mathbf{k}\mathbf{k}'}\cdot\mathbf{r}}
  \phi_\mu^{\mathbf{k}~*}(\mathbf{r})
  \phi_\nu^{\mathbf{k}'}(\mathbf{r}) d^3\mathbf{r}
  \nonumber \\
  &= \sum_{\mathbf{n}}e^{i\mathbf{k}'\cdot\mathbf{n}}
  \rho_{\mu^0\nu^\mathbf{n}}(\mathbf{G}_{\mathbf{k}\mathbf{k}'}),
  \label{eq:aft:rho} \\
  &= \sum_{\mathbf{m}}e^{-i\mathbf{k}\cdot\mathbf{m}}
  \rho_{\mu^\mathbf{m}\nu^0}(\mathbf{G}_{\mathbf{k}\mathbf{k}'})
  \label{eq:aft:rho:1} \\
  \mathbf{G}_{\mathbf{k}\mathbf{k}'} &= \mathbf{G}-\mathbf{k}+\mathbf{k}',
  \\
  \rho_{\mu^\mathbf{m}\nu^\mathbf{n}}(\mathbf{G})
  &= \int e^{-i\mathbf{G}\cdot\mathbf{r}}
  \mu(\mathbf{r}-\mathbf{m}) \nu(\mathbf{r}-\mathbf{n})d^3\mathbf{r};
  \label{eq:aft:rho:prim}
\end{align}
  \item A summation over PW basis inside the volume $\Omega$ of the unit cell
\begin{gather}
  (\phi_\mu^{\mathbf{k}} \phi_\nu^{\mathbf{k}'}|\phi_\kappa^{\mathbf{k}'} \phi_\lambda^{\mathbf{k}})
  = \frac{1}{\Omega}\sum_\mathbf{G}\mathcal{C}(\mathbf{G}_{\mathbf{k}\mathbf{k}'})
  \rho_{\mu\nu}^{\mathbf{k}\mathbf{k}'}(\mathbf{G})
  \rho_{\kappa\lambda}^{\mathbf{k}'\mathbf{k}}(-\mathbf{G}).
  \label{eq:eri:G}
\end{gather}
\end{enumerate}
$\mathcal{C}(\mathbf{G})$ in the equation above is the Coulomb kernel in reciprocal space.
The Coulomb metric $g(\mathbf{r}_{12})$ can be split into long-range (LR) plus
short-range (SR) contributions
\begin{equation}
  g(\mathbf{r}_{12}) = \frac{1}{r_{12}} = \frac{\erf(\omega r_{12})}{r_{12}} +
  \frac{\erfc(\omega r_{12})}{r_{12}}.
\end{equation}
When computing ERIs with Eq.~\eqref{eq:eri:G}, the Fourier transformed
full-range Coulomb kernel as well as the LR and SR Coulomb kernels are
\begin{gather}
  \frac{1}{r_{12}} \rightarrow
  \mathcal{C}(\mathbf{G}) =
  \begin{cases}
    \frac{4\pi}{G^2} & |\mathbf{G}| \neq 0 \\
    0                & |\mathbf{G}| =  0 
  \end{cases},
  \label{eq:coulomb:kernel}
  \\
  \frac{\erf(\omega r_{12})}{r_{12}} \rightarrow
  \mathcal{C}_\text{LR}(\mathbf{G}) =
  \begin{cases}
  \frac{4\pi}{G^2} \exp(-\frac{G^2}{4\omega^2}) & |\mathbf{G}| \neq 0 \\
  0                                           & |\mathbf{G}| =  0 
  \end{cases},
  \label{eq:lr:coulomb:kernel} \\
  \frac{\erfc(\omega r_{12})}{r_{12}} \rightarrow
  \mathcal{C}_\text{SR}(\mathbf{G}) = 
  \begin{cases}
  \frac{4\pi}{G^2} (1 - \exp(-\frac{G^2}{4\omega^2})) & |\mathbf{G}| \neq 0 \\
  \frac{\pi}{\omega^2}                              & |\mathbf{G}| =  0 
  \end{cases}
  \label{eq:sr:coulomb:kernel}.
\end{gather}
The treatment of $\mathbf{G} = 0$ in the SR Coulomb kernel is different to the other two.
When adding the LR and SR integrals to produce the integrals for full-range Coulomb,
an additional contribution corresponding to the FT term at $\mathbf{G} = 0$
needs to be subtracted from the SR integrals
\begin{equation}
  (\phi^{\mathbf{k}}_\mu\phi^{\mathbf{k}}_\nu|\phi^{\mathbf{k}}_\kappa\phi^{\mathbf{k}}_\lambda)^{\text{SR}}\Big|_{\mathbf{G}=0}
  = \frac{\pi}{\Omega\omega^2} S_{\mu\nu}^{\mathbf{k}} S_{\kappa\lambda}^{\mathbf{k}}.
  \label{eq:sr:eri:g0}
\end{equation}

Due to the exponential decay term $\exp(-\frac{G^2}{4\omega})$ in the LR Coulomb kernel,
we can expect that the contributions from large $\mathbf{G}$ in Eq.~\eqref{eq:eri:G} decrease rapidly.
The summation over PWs in Eq.~\eqref{eq:eri:G} can be truncated early thus a
small energy cutoff is required to compute LR ERI in reciprocal space.
Sparsity of SR ERIs in real space can be utilized to gain good efficiency\cite{Izmaylov2006}.
By choosing an appropriate value for $\omega$ and appropriate algorithms that we
will discuss below, computing SR ERIs in real space and LR ERIs in
reciprocal space appears to be an efficient approach.

In this approach, we use Ewald probe-charge correction\cite{Paier2005} to
regularize the Coulomb singularity in HFX.
It only requires to modify the FT integrals at $\mathbf{G} = 0$.
Other attractive techniques such as the Wigner-Seitz truncated Coulomb interactions\cite{Sundararaman2013}
are incompatible to the current range-separated algorithm.
They require more PWs to describe the Coulomb kernel while this algorithm only
use a small number of PWs for LR ERIs.

\subsection{HFX in reciprocal space}
\label{sec:kspace:ints}
Computing HFX in reciprocal space involves two time-consuming steps:
analytical FT for orbital product $\rho^{\mathbf{k}\mathbf{k}'}_{\mu\nu}(\mathbf{G})$
with $\mathcal{O}(N_k^2 N_c N_{AO}^2 N_G)$ complexity as shown by Eq.~\eqref{eq:aft:rho},
and the tensor contraction for three tensors which scales as $N_k^2 N_{AO}^3 N_G$
\begin{equation}
  K^{\mathbf{k}}_{\mu\lambda}
  = \frac{1}{N_k}\sum_{\nu\kappa\mathbf{G}\mathbf{k}'}
  D^{\mathbf{k}'}_{\nu\kappa} \rho^{\mathbf{k}\mathbf{k}'}_{\mu \nu}(\mathbf{G})
  \mathcal{C}(\mathbf{G}_{\mathbf{k}\mathbf{k}'})
  \rho^{\mathbf{k}'\mathbf{k}}_{\kappa \lambda}(-\mathbf{G}).
\end{equation}
The number of image cells in lattice-sum, $N_c$, is related to distance cutoff.
The number of PW basis $N_G$ needs to be determined by energy
cutoff. Distance cutoff and energy cutoff are influenced by the Coulomb
attenuation parameter $\omega$ and the compactness of the
Gaussian basis functions\cite{Sun2023}.
It should be noted that the FT for orbital products has to be evaluated
analytically. Using discrete FT here can cause significant errors\cite{Sun2017}.
This is because we only use a small number of PW basis for the
ERIs evaluated in reciprocal space. The FT associated real space grids are not
dense enough to describe all orbital products.

To optimize the performance of FT for orbital product, utilizing the symmetry
below can reduce the cost by a factor of two
\begin{equation}
  [\rho^{\mathbf{k}\mathbf{k}'}_{\mu\nu}(\mathbf{G})]^*
  = \rho^{\mathbf{k}'\mathbf{k}}_{\nu\mu}(-\mathbf{G})
  \label{eq:gpq:sym}.
\end{equation}
Thereby, the orbital product tensor can be computed once and used four times for
exchange matrices at two different $\mathbf{k}$ points
\begin{gather}
  K^{\mathbf{k}}_{\mu\lambda}
  = \sum_{\nu\kappa\mathbf{G}\mathbf{k}'}
  D^{\mathbf{k}'}_{\nu\kappa} \rho^{\mathbf{k}\mathbf{k}'}_{\mu \nu}(\mathbf{G})
  \mathcal{C}(\mathbf{G}_{\mathbf{k}\mathbf{k}'})
  [\rho^{\mathbf{k}\mathbf{k}'}_{\lambda\kappa}(\mathbf{G})]^*,
  \\
  K^{\mathbf{k}'}_{\nu\kappa}
  = \sum_{\mu\lambda\mathbf{G}\mathbf{k}}
  D^{\mathbf{k}}_{\mu\lambda} \rho^{\mathbf{k}'\mathbf{k}}_{\nu\mu}(-\mathbf{G})
  \mathcal{C}((-\mathbf{G})_{\mathbf{k}'\mathbf{k}})
  [\rho^{\mathbf{k}'\mathbf{k}}_{\kappa\lambda}(-\mathbf{G})]^*.
\end{gather}

During the self-consistent field (SCF) iteration, density matrices are
constructed from occupied SCF orbitals (denoted as
$C^{\mathbf{k}}_{\mu j}$ below) and orbital occupancies (denoted as
$\Lambda^{\mathbf{k}}_{j}$).
A common optimization practice is to use the
occupied orbitals directly in the construction of exchange matrix.
We can first carry out the contraction between the $N_{occ}$ occupied orbitals and the orbital product tensor
\begin{align}
  \rho^{\mathbf{k}\mathbf{k}'}_{\mu j}(\mathbf{G})
  &= \sum_\nu \rho^{\mathbf{k}\mathbf{k}'}_{\mu\nu}(\mathbf{G}) C^{\mathbf{k}'}_{\nu j},
  \label{eq:lpq:mo} \\
  \rho^{\mathbf{k}'\mathbf{k}}_{j \nu}(-\mathbf{G})
  &= \sum_\mu C^{\mathbf{k}'~*}_{\mu j} \rho^{\mathbf{k}'\mathbf{k}}_{\mu\nu}(-\mathbf{G})
  = \sum_\mu C^{\mathbf{k}'~*}_{\mu j} [\rho^{\mathbf{k}\mathbf{k}'}_{\nu\mu}(\mathbf{G})]^*
  = [\rho^{\mathbf{k}\mathbf{k}'}_{\nu j}(\mathbf{G})]^*,
\end{align}
then obtain the exchange matrix with a contraction between the resultant tensors
and another orbital product tensor
\begin{align}
  K^{\mathbf{k}}_{\mu\nu}
  &= \frac{1}{\Omega}\sum_{j\mathbf{G}\mathbf{k}'} \Lambda^{\mathbf{k}'}_{j}
  \rho^{\mathbf{k}\mathbf{k}'}_{\mu j}(\mathbf{G})
  \mathcal{C}(\mathbf{G}_{\mathbf{k}\mathbf{k}'}) \rho^{\mathbf{k}'\mathbf{k}}_{j \nu}(-\mathbf{G})
  \nonumber\\
  &= \frac{1}{\Omega}\sum_{j\mathbf{G}\mathbf{k}'} \Lambda^{\mathbf{k}'}_{j}
  \rho^{\mathbf{k}\mathbf{k}'}_{\mu j}(\mathbf{G})
  \mathcal{C}(\mathbf{G}_{\mathbf{k}\mathbf{k}'}) [\rho^{\mathbf{k}\mathbf{k}'}_{\nu j}(\mathbf{G})]^*.
\end{align}
This formula results in a reduction of the computational cost to
$N_k^2 N_{AO}^2 N_{occ} N_G$ for the step of tensor contraction.
The permutation symmetry Eq.~\eqref{eq:gpq:sym} for orbital product tensor can
still be utilized, leading to the contraction for the exchange matrix at
$\mathbf{k}'$
\begin{gather}
  \rho^{\mathbf{k}'\mathbf{k}}_{\nu j}(-\mathbf{G})
  = \sum_\mu [\rho^{\mathbf{k}\mathbf{k}'}_{\mu\nu}(\mathbf{G})]^* C^{\mathbf{k}}_{\mu j},
  \label{eq:lpq:mo1} \\
  K^{\mathbf{k}'}_{\mu\nu}
  = \frac{1}{\Omega}\sum_{j\mathbf{G}\mathbf{k}} \Lambda^{\mathbf{k}}_{j}
  \rho^{\mathbf{k}'\mathbf{k}}_{\mu j}(-\mathbf{G})
  \mathcal{C}( \mathbf{G}_{\mathbf{k}\mathbf{k}'})
  [\rho^{\mathbf{k}'\mathbf{k}}_{\nu j}(-\mathbf{G})]^*.
\end{gather}
By changing the contraction order in Eq.~\eqref{eq:lpq:mo}, we obtain
\begin{gather}
  \rho^{\mathbf{k}\mathbf{k}'}_{\mu j}(\mathbf{G})
  = \sum_{\nu\mathbf{n}} \rho_{\mu^0\nu^\mathbf{n}}(\mathbf{G}_{\mathbf{k}\mathbf{k}'})
  C_{\nu^\mathbf{n} j^{\mathbf{k}'}},
  \\
  C_{\nu^\mathbf{n} j^\mathbf{k}}
  = e^{i \mathbf{k}\cdot \mathbf{n}}C^{\mathbf{k}}_{\nu j}.
\end{gather}
It is noticeable that the sparsity in the FT orbital product tensor can be
exploited to reduce the computational cost.
The FT integral $\rho_{\mu^\mathbf{m}\nu^\mathbf{n}}(\mathbf{G}_{\mathbf{k}\mathbf{k}'})$ in
Eq.~\eqref{eq:aft:rho:prim} may be negligible for well-separated
$\mu(\mathbf{r-\mathbf{m}})$ and
$\nu(\mathbf{r}-\mathbf{n})$.
Although the tensor contraction still has the scaling $N_k^2 N_c N_{AO}^2 N_G$, the actual
cost is much lower due to the sparsity of
$\rho_{\mu^0\nu^\mathbf{n}}(\mathbf{G}_{\mathbf{k}\mathbf{k}'})$.

For SCF orbitals which preserve the time-reversal symmetry between $\mathbf{k}$
and $\bar{\mathbf{k}}$ ($=-\mathbf{k}$), it is possible to achieve an additional
reduction in computational cost.
The time-reversal symmetry ensures that the SCF wavefunction defined by the
$\mathbf{k}$-point orbitals can be transformed to
an identical representation at certain $\Gamma$-point with real orbital
coefficients $\mathbf{B}$
\begin{gather}
  B_{\mu^\mathbf{m} j^\mathbf{k}}
  = \frac{1}{\sqrt{2}}( C_{\mu^\mathbf{m} j^\mathbf{k}}
  + C_{\mu^\mathbf{m} j^{\bar{\mathbf{k}}}}),
  \\
  B_{\mu^\mathbf{m} j^{\bar{\mathbf{k}}}}
  = \frac{1}{\sqrt{2}}( C_{\mu^\mathbf{m} j^\mathbf{k}}
  - C_{\mu^\mathbf{m} j^{\bar{\mathbf{k}}}}).
\end{gather}
This property reduces the cost of comupting
$\rho^{\mathbf{k}\mathbf{k}'}_{\mu j}(\mathbf{G})$ by roughly a factor of two as we only
need to compute the contraction between a complex tensor and a real matrix
\begin{gather}
  L^{\mathbf{k}}_{\mu j}(\mathbf{G})
  = \sum_{\nu\mathbf{n}} \rho_{\mu^0\nu^\mathbf{n}}(\mathbf{G}) B_{\nu^\mathbf{n} j^{\mathbf{k}}},
  \label{eq:lpj:a} \\
  \rho^{\mathbf{k}\mathbf{k}'}_{\mu j}(\mathbf{G})
  = \frac{1}{\sqrt{2}}L^{\mathbf{k}'}_{\mu j}(\mathbf{G}_{\mathbf{k}\mathbf{k}'})
  - \frac{i}{\sqrt{2}}L^{\bar{\mathbf{k}}'}_{\mu j}(\mathbf{G}_{\mathbf{k}\mathbf{k}'}).
\end{gather}
In addition, the intermediates $L^{\mathbf{k}}_{\mu j}$ can be reused
for the tensor contraction in Eq.~\eqref{eq:lpq:mo1},
providing another factor of two reduction in computational cost.
We first employ the FT Eq.~\eqref{eq:aft:rho:1}
\begin{gather}
  \rho^{\mathbf{k}'\mathbf{k}}_{\nu j}(-\mathbf{G})
  = \sum_{\mu\mathbf{m}}[\rho_{\mu^\mathbf{m}\nu^0}(\mathbf{G}_{\mathbf{k}\mathbf{k}'})
  e^{-i\mathbf{k}\cdot \mathbf{m}}]^* C^\mathbf{k}_{\mu j}
  = \sum_{\mu\mathbf{m}}[\rho_{\nu^0\mu^\mathbf{m}}(\mathbf{G}_{\mathbf{k}\mathbf{k}'})
  C_{\mu^\mathbf{m}j^\mathbf{k}}^*]^*.
\end{gather}
Realizing that the orbital coefficients $\mathbf{B}$ are real,
\begin{gather}
  C_{\mu^\mathbf{m}j^\mathbf{k}}^*
  = \frac{1}{\sqrt{2}}(B_{\mu^\mathbf{m}j^\mathbf{k}} + i B_{\mu^\mathbf{m}j^\mathbf{k}})
\end{gather}
we then simplify Eq.~\eqref{eq:lpq:mo1} to
\begin{gather}
  \rho^{\mathbf{k}'\mathbf{k}}_{\nu j}(-\mathbf{G})
  = \frac{1}{\sqrt{2}}[L^{\mathbf{k}}_{\nu j}(\mathbf{G}_{\mathbf{k}\mathbf{k}'})]^*
  - \frac{i}{\sqrt{2}}[L^{\bar{\mathbf{k}}}_{\nu j}(\mathbf{G}_{\mathbf{k}\mathbf{k}'})]^*.
\end{gather}
By using all the features discussed above, we would obtain an overall
reduction which rougly equals a factor of 8 plus speed-ups from tensor sparsity.
Finally, we derive the exchange matrices 
\begin{gather}
  K^{\mathbf{k}}_{\mu\nu}
  = \frac{1}{2}\sum_{\mathbf{k}'}
  ( \tilde{K}^{\mathbf{k}'\mathbf{k}'}_{\mu\nu}
  +i\tilde{K}^{\mathbf{k}'\bar{\mathbf{k}}'}_{\mu\nu}
  -i\tilde{K}^{\bar{\mathbf{k}}'\mathbf{k}'}_{\mu\nu}
  + \tilde{K}^{\bar{\mathbf{k}}'\bar{\mathbf{k}}'}_{\mu\nu}),
  \\
  K^{\mathbf{k}'}_{\mu\nu}
  = \frac{1}{2}\sum_{\mathbf{k}}
  ( \tilde{K}^{\mathbf{k}\mathbf{k}}_{\nu\mu}
  +i\tilde{K}^{\mathbf{k}\bar{\mathbf{k}}}_{\nu\mu}
  -i\tilde{K}^{\bar{\mathbf{k}}\mathbf{k}}_{\nu\mu}
  + \tilde{K}^{\bar{\mathbf{k}}\bar{\mathbf{k}}}_{\nu\mu}),
\end{gather}
where $\tilde{K}$ is the intermediate based on $L^{\mathbf{k}}_{\mu j}$
\begin{equation}
  \tilde{K}^{\mathbf{k}_1\mathbf{k}_2}_{\mu\nu}
  =\frac{1}{N_k}\frac{1}{\Omega}\sum_{j\mathbf{G}}\Lambda^{\mathbf{k}_1}_i
  L^{\mathbf{k}_1}_{\mu j}(\mathbf{G}_{\mathbf{k}\mathbf{k}'})
  \mathcal{C}(\mathbf{G}_{\mathbf{k}\mathbf{k}'})
  [L^{\mathbf{k}_2}_{\nu j}(\mathbf{G}_{\mathbf{k}\mathbf{k}'})]^*.
\end{equation}
For sufficient $\mathbf{k}$ points, this intermediate can approximately lower
25\% of the cost due to the relation
\begin{equation}
  \tilde{K}^{\bar{\mathbf{k}}\mathbf{k}}_{\mu\nu}
  = [\tilde{K}^{\mathbf{k}\bar{\mathbf{k}}}_{\nu\mu}]^*.
\end{equation}

\subsection{SR integrals evaluation}
\label{sec:sr:eri}

Following to the idea proposed in \onlinecite{Irmler2018},
we first evaluate SR Coulomb integrals as well as the Coulomb and exchange
matrix in the Born-von Karman (BvK) supercell then transform them to the
$\mathbf{k}$-point adapted quantities. For example,
the matrix element of SR HFX $K_{\mu\nu}^{\text{SR},\mathbf{k}}$ can be
computed
\begin{gather}
  K_{\mu^0\lambda^{\mathbf{l}}}^\mathrm{SR}
  = \sum_{\nu\kappa} \sum_{\mathbf{L}} \sum_{\mathbf{m}+\mathbf{M}} \sum_{\mathbf{n}+\mathbf{N}}
  D_{\nu^{\mathbf{n}}\kappa^{\mathbf{m}}}
  (\mu^{\mathbf{0}} \nu^{\mathbf{n}+\mathbf{N}}|\kappa^{\mathbf{m}+\mathbf{M}}
  \lambda^{\mathbf{l}+\mathbf{L}+\mathbf{M}})^\mathrm{SR},
  \label{eq:K:bvkcell}
  \\
  K_{\mu\nu}^{\text{SR},\mathbf{k}}
  = \sum_{\mathbf{l}} e^{i\mathbf{k}\cdot\mathbf{l}} K^\text{SR}_{\mu^0\nu^\mathbf{l}}.
\end{gather}
Here the lattice summation involves two types of translation vectors, the super-lattice translation
vectors ($\mathbf{L}, \mathbf{M}, \mathbf{N}$) for BvK supercell and the
translation inside a single BvK supercell ($\mathbf{l}, \mathbf{m}, \mathbf{n}$).
$D_{\nu^{\mathbf{n}}\kappa^{\mathbf{m}}}$ in Eq.~\eqref{eq:K:bvkcell}
represents the density matrix elements in the BvK supercell.
Considering the BvK supercell being filled with $N_\mathrm{BvK}$ primitive cells,
$D_{\nu^{\mathbf{l}}\kappa^{\mathbf{m}}}$ is obtained via an inverse FT of the
$\mathbf{k}$-point adapted density matrices
\begin{equation}
  D_{\nu^{\mathbf{n}}\kappa^{\mathbf{m}}}
  = D_{\nu^{\mathbf{0}}\kappa^{\mathbf{m}-\mathbf{n}}}
  = \frac{1}{N_\mathrm{BvK}}\sum_{\mathbf{k}}e^{-i\mathbf{k}\cdot(\mathbf{m}-\mathbf{n})} D_{\nu\kappa}^{\mathbf{k}}.
\end{equation}

The cost of computing SR integrals in real space mainly depends on the range of
lattice-sum and integral screening of the SR integrals. The number of
$\mathbf{k}$ points has relatively low impact on the computational cost.
The value of Coulomb attenuation parameter $\omega$
and the compactness of Gaussian basis functions are the two main factors
to determine the distance cutoff (as well as the range of
lattice-sum)\cite{Sun2023}.
Small value of $\omega$ and diffused basis functions can lead to more image
cells in lattice-sum.
Although the size of BvK super-cell is proportional to the size of $\mathbf{k}$
points, the overall number of image cells in the lattice-sum, denoted by the
compound index $\mathbf{m}+\mathbf{M}$, remains largely unchanged.

To further reduce the cost of SR integrals, we also take the locality of basis
functions into account.
The Gaussian basis functions can be partitioned into two groups, labelled as $D$ for
the diffused functions and $C$ for the compact functions.
The products of two functions $\rho_{\mu\nu}$ have four possible
combinations: the product of two diffused functions can lead to a diffuse
density $\rho_D$ while the rest three $DC$, $CD$ and $CC$ all lead to compact
densities $\rho_C$. Here we developed three recipes to evaluate these SR
integrals. Each recipe is adequate in specific scenarios, depending on the size of
$\mathbf{k}$-point mesh.

\begin{itemize}
  \item Recipe I: Analytically evaluating the 9 types of SR integrals which only
    consist of compact densities $(DC|DC)$, $(DC|CD)$,
    $(DC|CC)$, $(CD|DC)$, $(CD|CD)$, $(CD|CC)$, $(CC|DC)$, $(CC|CD)$, $(CC|CC)$
    in real space while leaving the remaining 7 types of SR integrals
    to the reciprocal space prescription we presented in Section \ref{sec:kspace:ints}.
    Since the 9 types of analytical integrals involve only compact density
    distributions, lattice-sum can be rapidly truncated.

    For the integrals involving diffuse densities, for instance the
    $(CC|DD)$-type integrals, the computation of SR integrals and their LR
    counterparts can be merged into one pass for $(CC|DD)$ with the full-range
    Coulomb kernel since both the SR and LR integrals are evaluated in reciprocal
    space. It is important to choose an
    appropriate energy cutoff $E_{cut}$ to ensure the accuracy of the
    $(CC|DD)$-type full-range ERIs. Given the precision requirement $\epsilon$, the energy
    cutoff for four-center ERIs can be estimated\cite{Sun2023}
    \begin{gather}
      16\pi^2 \theta E_{cut}^{\frac{1}{2}}
      (\frac{E_{cut}}{2\alpha_C})^{2l_C} (\frac{E_{cut}}{2\alpha_D})^{2l_D}
      e^{\frac{E_{cut}}{2\theta}} < \epsilon,
      \label{eq:alpha:cut} \\
      \theta = \frac{1}{2}(\alpha_C^{-1} + \alpha_D^{-1})^{-1},
    \end{gather}
    where $\alpha_C$ and $l_C$ are the exponent and the angular moment of the most
    compact basis function. We can solve the inequality \eqref{eq:alpha:cut} for
    each diffused function and find an overall energy cutoff for the reciprocal
    space prescription.

    When computing exchange matrix with this recipe, it is inconvenient to
    merely compute the 7 types of diffuse-density associated SR integrals with the
    reciprocal space prescription.
    We instead use the reciprocal space prescription to compute the HFX of all
    integrals as well as the SR HFX of the 9 types of compact SR integrals.
    Then, we take the difference between the two results
    to obtain the contribution of the 7-type SR integrals.

    This recipe lowers the cost of analytical integral evaluation while it
    introduces an extra HFX computation in reciprocal space.
    The cost for the later one is proportional to $N_k^2$ and it can dominate
    the HFX computation if the SCF calculation involves many
    $\mathbf{k}$ points. Therefore, this recipe is only adequate for SCF
    calculations with a small number of $\mathbf{k}$ points.

  \item Recipe II: Computing the $(DD|DD)$-type SR integral with
    the reciprocal space prescription while handling the rest 15 types in real
    space. $(DD|DD)$-type SR integrals in real space often involves a large
    number of image cells in the triple lattice-sum as it requires the largest
    distance cutoff. Excluding it from the real space integral evaluation can
    save computational cost.
    Meanwhile, evaluating $(DD|DD)$-type SR integrals in reciprocal space only adds a
    small overhead, as the diffused functions typically constitute a small
    portion of the regular basis set.
    This algorithm is helpful in reducing the computation time of Fock build for
    a moderate number of $\mathbf{k}$ points.

  \item Recipe III: Evaluating all SR integrals in real space.
    When a large number of $\mathbf{k}$ points is present in
    the SCF calculation, the effort of computing exchange matrix in reciprocal
    space become dominant.
    To reduce the cost of reciprocal space part, we
    not only evaluate all SR integrals in real space, but also tend to use
    small energy cutoff to decrease the required number of PWs which is proportional
    to the computational cost of reciprocal space algorithm.
\end{itemize}

The value of Coulomb attenuation parameter $\omega$, energy cutoff, and the
number of PWs are correlated.
Although less PWs means less computational efforts in reciprocal space
part, a lower energy cutoff can lead to more computation
time in real space as it requires a small $\omega$ to ensure the integral
accuracy.
To gain the optimal performance, different $\omega$ should be used in different
$\mathbf{k}$-point calculations.

\subsection{Integral screening}

Integral screening is a crucial factor for the performance of real space
integral program. To enhance the performance of SR integral computation, we have
developed a three-layer hierarchical screening prescription:
\begin{itemize}
\item First, we filter integrals with the regular Schwarz inequality
\begin{gather}
  (\mu\nu|\kappa\lambda)^\text{SR}
  < \sqrt{(\mu\nu|\mu\nu)^\text{SR} (\kappa\lambda|\kappa\lambda)^\text{SR}}.
\end{gather}

\item Next, we apply the mutated Schwarz inequality to estimate SR integral value
\begin{gather}
  (\mu\nu|\kappa\lambda)^\text{SR} <
  \min(\sqrt{(\mu\mu|\kappa\kappa)^\text{SR} (\nu\nu|\lambda\lambda)^\text{SR}},
  \sqrt{(\mu\mu|\lambda\lambda)^\text{SR} (\nu\nu|\kappa\kappa)^\text{SR}}).
\end{gather}
This condition is not useful for ERIs of full Coulomb metric due to the
long range character of Coulomb interactions.
However, for SR integrals, this condition can filter out approximately 50\% of
the terms.

\item Finally, the SR integral estimator developed in \onlinecite{Sun2023} is
  utilized as the last integral filter
\begin{equation}
  (\mu\nu|\kappa\lambda)^\text{SR}
  \lesssim \frac{Q_{\mu\nu}Q_{\kappa\lambda}e^{-\theta_{\mu\nu\kappa\lambda}R^2}}{R^2}
\end{equation}
where $Q_{\mu\nu}$ is a factor that behaves like the overlap between two primitive
Gaussian functions. Its detailed definition can be found in \onlinecite{Sun2023}.
Other quantities can be computed
\begin{gather}
  \theta_{\mu\nu\kappa\lambda}
  = ((\alpha_\mu + \alpha_\nu)^{-1} + (\alpha_\kappa + \alpha_\lambda)^{-1} + \omega^{-2})^{-1},
  \\
  R = |\mathbf{P}_{\mu\nu} - \mathbf{P}_{\kappa\lambda}|,
  \\
  \mathbf{P}_{\mu\nu}
  = \frac{\alpha_\mu \mathbf{R}_\mu + \alpha_\nu \mathbf{R}_\nu}{\alpha_\mu + \alpha_\nu}.
\end{gather}
\end{itemize}

The reason that the screening filters are applied in this specific order is that each
screening filter has a different computational overhead.
We have observed that the time spent on the screening testing is not
negligible due to the large number of integral quartets that needs to be tested.
The regular Schwarz inequality is applied first, as it is not only
effective in filtering out most small integrals but also the fastest.
The regular Schwarz inequality test requires only
one multiplication and two memory accesses. The mutated Schwarz inequality is
slightly less efficient as it requires two multiplications and four memory accesses.
The integral estimator works slowest.
Even though some intermediates are precomputed and cached,
it requires six memory accesses to obtain coordinates for $R$, two
memory accesses for paired quantities $Q$, one call to the exponential function, and ten
or more multiplication/addition/division operations to aggregate them.

\section{\label{sec:results}Numerical tests}

The range-separated Fock build algorithm has been implemented in the PySCF program
package\cite{Sun2020}.
In all exhibited calculations,
the precision requirement for energy cutoff and distance cutoff was set to $10^{-8}$.
Timing is measured on a desktop computer with Intel I9-7940X 14-core CPU and 64
GB memory.

Our first tests are all-electron k-point Hartree-Fock calculations for diamond
crystal. Gaussian basis sets STO-3G and cc-pVDZ are employed in our tests.
System size ranges from $1\times 1\times 1$ $\mathbf{k}$ point ($\Gamma$ point) to
$10\times 10\times 10$ $\mathbf{k}$ points.
The results are collected in Table \ref{tab:sto3g} and \ref{tab:vdz}.
Using the range-separated algorithm introduced here, we can accomplish the
calculation of cc-pVDZ basis with $4\times 4\times 4$ $\mathbf{k}$-point HF in
20 CPU hours.  Even for a dense $10\times 10\times 10$ $\mathbf{k}$-mesh, which
consists of 8000 atoms and 112000 basis functions, the SCF calculation takes
only 254 CPU hours with a memory footprint less than 6 GB.

The number of PWs and the corresponding range-separation parameter
$\omega$ in our $\mathbf{k}$-point HF calculations are also listed in Table
\ref{tab:sto3g} and \ref{tab:vdz}.
For the small $\mathbf{k}$ points, it is often less demanding to compute LR HFX in
reciprocal space. We tend to use more PWs to increase the
value of $\omega$ for SR HFX because large $\omega$ can help SR Coulomb
potential rapidly decay in real space.
In this scenario, the associated energy cutoff is high enough for the full-range
Coulomb integrals with some diffused basis.
We then use this criteria to partition basis functions and employ the SR
integral recipe \Romannum{1} and recipe \Romannum{2} to
reduce the cost of the real-space SR integral evaluation.
As mentioned in Section \ref{sec:sr:eri}, the reciprocal space HFX algorithm
needs to be executed twice when applying recipe \Romannum{1}.
This overhead is noticeable in the $4\times 4\times 4$ $\mathbf{k}$-point
calculation with cc-pVDZ basis.
With the number of PWs chosen in Table \ref{tab:vdz},
the ratio of theoretical scaling between the LR part of the $4\times 4\times 4$ mesh and
the $3\times 3\times 3$ mesh is $64^2 \times 13^3 / (27^2 \times 17^3) \approx 2.5$.
By using recipe \Romannum{2} for SR HFX, the actual resources spent on the LR part of the
$4\times 4\times 4$ mesh is 6.5 CPU hours, which is only 70\% more than the
resources for the $3\times 3\times 3$ mesh.

The computation cost of HFX in the reciprocal space part increases rapidly as
we increase the number of $\mathbf{k}$ points, while the real space SR HFX has a
relatively slower increase in cost.
Therefore, we prefer SR HFX recipe \Romannum{3} with fewer PWs
for calculations with a large number of $\mathbf{k}$ points.
The sparsity in orbital products exhibits effectiveness for large
$\mathbf{k}$-mesh calculations.
For example, the average time of Fock build spent on the LR HFX of $10\times 10\times 10$
$\mathbf{k}$-point calculation is 13 CPU hours, which is 25\%
more than that of the $8\times 8\times 8$ case,
although the theoretical scaling indicates 40\% more FLOPs for the former.

The help of sparsity is more effective for calculations with denser $\mathbf{k}$-point mesh.
This enables us to perform all-electron Hartree-Fock calculations for
LiH crystal up to $24\times 24\times 24 = 13824$ $\mathbf{k}$-points, which amounts to a
total of 27648 atoms and 1036800 basis functions. The largest $\mathbf{k}$-point HF
calculation for LiH crystal takes 1380 CPU hours in which 1323 CPU hours are
spent on the LR HFX computation. In this calculation, we use a small value of $\omega = 0.14$
for SR HFX which results in 27 PWs for LR HFX.

Large scale calculations towards thermodynamic limit of HF methods with Gaussian
basis for LiH crystal were studied in \onlinecite{Joachim2009}. In that work,
two methods were
employed to obtain the HF energy: an extrapolation from SR HFX to full-range
HFX with Pad\'e approximants, and a direct evaluation of HFX with
truncated Coulomb interactions.
Our $\mathbf{k}$-point HF results are listed in Table \ref{tab:lih}. 
As we increase the $\mathbf{k}$-point mesh, the obtained HF energy per unit cell
gradually converges towards -8.064539 $E_h$, which is 4 $\mu E_h$
higher than the extrapolation result or 6 $\mu E_h$ higher than the direct
evaluation result reported in \onlinecite{Joachim2009}.
Based on the trends we observed in our calculation series, we can expect that
the HF thermodynamic limit is probably several $\mu E_h$ higher than the result we obtained.
The difference may be attributed to two possibilities.
The first is the treatment of Coulomb singularity in HFX.
We utilize the full-range Coulomb interactions with Ewald probe charge correction in
this work. In contrast, the extrapolation method does not have any
treatment for singularity in $\mathbf{k}$ sampling while the result of direct
evaluation method is based on manually tuned truncated Coulomb potentials.
The second possibility is the size of $\mathbf{k}$-point mesh utilized in the
calculation.
The calculation of direct evaluation method includes only 1000 atoms in a
$5\times 5\times 5$ mesh of the cubic cell structure which is smaller than the
system size in our tests.

\begin{table*}
\caption{Results of all-electron $\mathbf{k}$-point HF with STO-3G basis for diamond crystal.
The unit cell is a cubic cell with lattice parameter 3.5668 \AA.
HF energy is given in Hartree atomic units.}
\label{tab:sto3g}
\begin{ruledtabular}
  \begin{tabular}{llcccccc}
    $\mathbf{k}$ mesh & $E_\text{HF}$ per cell & timing/h & Fock builds
    & $\omega$ & $N_G$ & SR-ERI recipe & SR-ERI timing \\
    \hline
$1 \times 1 \times 1 $ & -299.328101 & 0.14 & 9 & 1.15 & $23^3$ & \Romannum{1} & 0.04 \\
$2 \times 2 \times 2 $ & -299.551274 & 0.64 & 8 & 0.94 & $19^3$ & \Romannum{1} & 0.23 \\
$3 \times 3 \times 3 $ & -299.525890 & 0.72 & 8 & 0.73 & $15^3$ & \Romannum{1} & 0.30 \\
$4 \times 4 \times 4 $ & -299.516150 & 1.09 & 8 & 0.52 & $11^3$ & \Romannum{1} & 0.41 \\
$6 \times 6 \times 6 $ & -299.510660 & 5.48 & 9 & 0.42 & $9 ^3$ & \Romannum{3} & 2.01 \\
$8 \times 8 \times 8 $ & -299.509296 & 13.9 & 9 & 0.31 & $7 ^3$ & \Romannum{3} & 3.70 \\
$10\times 10\times 10$ & -299.508809 & 22.7 & 9 & 0.21 & $5 ^3$ & \Romannum{3} & 8.58 \\
  \end{tabular}
\end{ruledtabular}
\end{table*}

\begin{table*}
\caption{Results of all-electron $\mathbf{k}$-point HF with cc-pVDZ basis for diamond crystal.
The unit cell is a cubic cell with lattice parameter 3.5668 \AA.
HF energy is given in Hartree atomic units.}
\label{tab:vdz}
\begin{ruledtabular}
  \begin{tabular}{llcccccc}
    $\mathbf{k}$ mesh & $E_\text{HF}$ per cell & timing/h & Fock builds
    & $\omega$ & $N_G$ & SR-ERI recipe & SR-ERI timing \\
    \hline
$1 \times 1 \times 1 $ & -302.870240 &  0.58 & 9 & 1.15 & $23^3$ & \Romannum{1} &  0.30 \\
$2 \times 2 \times 2 $ & -303.076048 &  2.71 & 8 & 1.04 & $21^3$ & \Romannum{1} &  0.62 \\
$3 \times 3 \times 3 $ & -303.054159 &  8.23 & 9 & 0.83 & $17^3$ & \Romannum{1} &  4.41 \\
$4 \times 4 \times 4 $ & -303.044553 &  19.7 & 9 & 0.63 & $13^3$ & \Romannum{2} &  13.2 \\
$6 \times 6 \times 6 $ & -303.038978 &  54.3 & 9 & 0.42 & $9 ^3$ & \Romannum{2} &  25.8 \\
$8 \times 8 \times 8 $ & -303.037585 & 134.7 & 9 & 0.31 & $7 ^3$ & \Romannum{3} &  40.7 \\
$10\times 10\times 10$ & -303.037088 & 254.5 & 11& 0.21 & $5 ^3$ & \Romannum{3} &  110.0 \\
  \end{tabular}
\end{ruledtabular}
\end{table*}

\begin{table*}
\caption{$\mathbf{k}$-point HF tests for LiH crystal at experimental lattice constant 4.084\AA.}
\label{tab:lih}
\begin{ruledtabular}
  \begin{tabular}{llcccccc}
    $\mathbf{k}$ mesh & $E_\text{HF}$ per cell \\
$1 \times 1 \times 1 $ &   -8.449722 \\
$2 \times 2 \times 2 $ &   -8.063001 \\
$3 \times 3 \times 3 $ &   -8.072053 \\
$4 \times 4 \times 4 $ &   -8.066987 \\
$5 \times 5 \times 5 $ &   -8.065904 \\
$6 \times 6 \times 6 $ &   -8.065308 \\
$7 \times 7 \times 7 $ &   -8.065022 \\
$8 \times 8 \times 8 $ &   -8.064859 \\
$9 \times 9 \times 9 $ &   -8.064760 \\
$11\times 11\times 11$ &   -8.064654 \\
$13\times 13\times 13$ &   -8.064603 \\
$15\times 15\times 15$ &   -8.064577 \\
$19\times 19\times 19$ &   -8.064551 \\
$24\times 24\times 24$ &   -8.064539 \\
  \end{tabular}
\end{ruledtabular}
\end{table*}

\section{Conclusions}
In this work, we presented an efficient range-separated algorithm to compute
electron repulsion integrals and Fock matrix for crystalline Gaussian basis.
In this algorithm, the full-range Coulomb interactions were split into the
short-range and long-range parts, which are respectively computed in real and
reciprocal space.
Additionally, we proposed several methods to improve the computational efficiency.
For integrals evaluated in reciprocal space,
we reformulated the Fourier transform and the tensor contraction equations
to utilize sparsity in Gaussian orbital products.
For short-range integrals,
we partitioned basis functions according to their locality in real space and
developed three recipes to utilize the localities of basis functions.
Different recipes are adequate for different scenarios depending on the
number of $\mathbf{k}$ points.
We developed a three-level integral screening scheme to filter SR electron
repulsion integrals which has significant impacts on the computational
efficiency for short-range integrals.
This algorithm enables us to perform all-electron HF and hybrid functional
DFT calculations with a large number of $\mathbf{k}$ points.
To assess the performance of this algorithm, we performed $\mathbf{k}$-point HF
calculations for LiH crystal with up to 1m basis functions.
The current algorithm only supports the Ewald probe-charge correction in the
treatment of Coulomb singularity.
Combining this algorithm with other Coulomb singularity treatments will be
considered in the future.

\clearpage
\bibliography{range_separation_jk}

\end{document}